# COSMIC RAY ANISOTROPY STUDY BY MEANS OF DETECTION OF MUON BUNDLES

G. Trinchero[1], M.B. Amelchakov[2], A.G. Bogdanov[2], A. Chiavassa[2,3,4], A.N. Dmitrieva[2], G. Mannocchi[1], S.S. Khokhlov[2], R.P. Kokoulin[2], K.G. Kompaniets[2], A.A. Petrukhin[2], V.V. Shutenko[2], I.A. Shulzhenko[2], I.I. Yashin[2], E.A. Yurina[2]

ABSTRACT

In this work, we use muon bundles which are formed in extensive air showers and detected at the ground level as a tool for searching anisotropy of high energy cosmic rays. Such choice is explained by the penetrating ability of muons which allows them to retain the direction of primary particles with a good accuracy. In 2012-2022, we performed long-term muon bundle detection with the coordinate-tracking detector DECOR, which is a part of the Experimental complex NEVOD (MEPhI, Moscow). To search for the cosmic rays anisotropy, the muon bundles arriving at zenith angles in the range from 15° to 75° in the local coordinate system are used. During the entire period of data taking, about 14 million of such events have been accumulated. In this paper, we describe some methods developed in the Experimental complex NEVOD and implemented in our research, including: the method for compensating the influence of meteorological conditions on the intensity of muon bundles at the Earth's surface, the method for accounting the design features of the detector and the inhomogeneity of the detection efficiency for different directions, as well as the method for estimating primary energies of cosmic rays. Here we present the results of the search for the dipole anisotropy of cosmic rays with energies in the PeV region and also compare them with the results obtained at the other scientific facilities.

## 1. INTRODUCTION

Cosmic ray (CR) anisotropy is usually defined as the relative deviation from the assumed isotropic flux. Study of CR anisotropy on the Earth's surface is a very complicated problem which holds a lot of puzzles. The phenomenon of cosmic ray anisotropy is associated with the distribution of the CR sources on the celestial sphere, as well as with the interaction of CR with matter and fields in the interstellar space. The so-called small scale anisotropy (SSA) is associated with local sources. Thus, on the celestial sphere the perturbation of the CR flux has a local character. To observe the SSA, it is necessary that the particles emitted from the source had rather high energy to retain the direction during their passing through the scattered magnetic

---

[1] Osservatorio Astrofisico di Torino – INAF, Italy
[2] National Research Nuclear University MEPhI (Moscow Engineering Physics Institute), Moscow, 115409 Russia
[3] Dipartimento di Fisica dell' Università degli Studi di Torino, 10125 Torino, Italy
[4] Sezione di Torino dell' Istituto Nazionale di Fisica Nucleare – INFN, 10125 Torino, Italy

fields in the Galaxy. Typical interstellar magnetic field in the Galaxy is about 3 μG (Malcolm 2011). For example, the proton with energy of 1 PeV has a gyroradius of about 3 light years in the Galaxy. It corresponds to the distances to the stars closest to the Sun (4 - 5 light years). To observe SSA from more distant objects, the energy of the emitted charged particles must be much higher.

An uneven distribution of CR sources and magnetic fields in the Galaxy create conditions for CR diffusion which, in its turn, could cause the large scale anisotropy (LSA). Since the Solar system is on the periphery of our Galaxy, we should expect an excess of CR from the central part of the Galaxy. But also the Sun is located at the inner edge of the Orion Spur (Reid & Zheng 2020) which could be the other reason of LSA. Existing models of CR diffusion from the center of the Galaxy predict that the amplitude of the dipole anisotropy depends on the rigidity of primary particles as ~ $R^{1/3}$ (Berezinsky 1990).

The other reason of LSA is Compton-Getting effect (Compton & Getting 1935). It occurs when the detector moves relative to the flux of cosmic rays. In this case, the anisotropy can be expressed by the formula (Gleeson & Axford 1968):

$$A = (\gamma + 2)\frac{w}{v}. \qquad (1)$$

where $\gamma$ is the absolute value of the power-law index of the CR differential energy spectrum, $w$ is the detector velocity relative to the cosmic ray flux, $v$ is the particle velocity.

Since the detector is located on the Earth's surface, it is possible to consider the orbital speed of the Earth (~ 30 km/s) around the Sun and the peculiar motion of the Sun (~ 20 km/s) with respect to the so-called local standard of rest (Schönrich et al. 2010). In these cases the calculated amplitude of anisotropy is of about $10^{-4}$. A circular rotation speed of the Sun around the center of the Galaxy is 240 ± 8 km/s (Reid et al 2014), and the amplitude of anisotropy should be by an order of magnitude greater (~ $10^{-3}$), but only for the extragalactic flux of CRs with energy higher than 1000 PeV.

In ground-based installations, the components of extensive air showers (EAS) are used as tools to study anisotropy. The EAS components appear as a result of the development of nuclear-electromagnetic cascades in the atmosphere. Therefore, a change in the state of the atmosphere affects the development of cascades. To correct data for the atmospheric effect in the analysis, the East-West method (Bonino et al 2011) is used in most experiments. The total counting rates of events observed in either the Eastern or the Western half of the field of view of installations can be varied due to different factors during a sidereal day that may complicate the search for anisotropy. The East-West method is aimed at reconstructing the equatorial component of a genuine large scale pattern of anisotropy by using only the difference of the counting rates from the Eastern and Western hemispheres.

## 2. DIPOLE ANISOTROPY SEARCH

Isotropy means that some object or phenomenon has the same physical properties in all directions. In its turn, the term "anisotropy" is used to describe situations where properties vary systematically, dependent on direction. In case of dipole anisotropy the resulting CR flux would be described by the formula:

$$I = I_0\left(1 + \vec{r}\vec{D}\right), \qquad (2)$$

where $I_0$ is the isotropic flux, $\vec{r}$ is the unit vector of observation direction, $\vec{D}$ is the dipole vector. The flux $I$ has the maximum value when the observation vector and the anisotropy dipole have the same direction. Consequently, the amplitude of anisotropy of the cosmic ray flux for selected angle of view ($\psi$) relative to the dipole direction can be expressed as

$$A = d\cos\psi = \frac{I - I_0}{I_0}, \qquad (3)$$

where $d$ is the module of $\vec{D}$, $\psi$ is the angle between vectors $\vec{D}$ and $\vec{r}$.

Determination of the dipole anisotropy parameters is not a completely trivial task. The problem is related to the spherical shape of the Earth. The installation located on its surface cannot observe the entire celestial sphere. In the 2$^{nd}$ equatorial coordinate system (Sadler et al. 1974) the formula (3) transforms to the following form:

$$\frac{\Delta I(\alpha,\delta)}{I_0} = d(\sin\delta\cos\delta_0 + \cos\delta\cos\delta_0\cos(\alpha - \alpha_0)), \qquad (4)$$

where $\alpha_0$ (right ascension) and $\delta_0$ (declination) are the coordinates of dipole anisotropy vector, $\alpha$ and $\delta$ define the direction of detection. In formula (4), the measured value of the relative deviation is associated with three unknown parameters ($d$, $\alpha_0$, $\delta_0$). To simplify the analysis we can use the projection of the anisotropy vector onto the plane. In this case, the most convenient plane is the equatorial one. Usually, the equatorial plane is used in order to provide the possibility of comparing the results obtained at the installations located in the southern and northern hemispheres. In the 2$^{nd}$ equatorial system, the projection on the equatorial plane is equivalent to projection onto the right ascension axis. To obtain the result, it is necessary to integrate formula (4) over the declination ($\delta$). In this case, the dipole anisotropy vector declination ($\delta_0$) remains undefined, as well as the absolute value of the vector itself. The only parameter determined from direct measurements is $\alpha_0$ which is called the anisotropy phase.

Attempts to study the anisotropy are made with almost all installations for detecting EAS components in a wide range of primary energies from TeV to EeV. The anisotropy amplitude in the projection on the right ascension axis varies in the range from $10^{-4}$ to $10^{-3}$. The region of primary energies near 200 TeV is of particular interest. Data from the ARGO-ABJ (Bartoli et al 2012), Tibet-AS$\gamma$ (Amenomori et al 2017), LHAASO-KM2A (Gao et al 2021) and IceCube (Aartsen et al 2016) facilities, which use different types of detectors and measure different EAS components, indicate that in this region a sharp change of the anisotropy dipole occurs in the observed direction and the anisotropy amplitude reaches its minimum. This phenomenon occurs just below the first "knee" in the energy spectrum, and its cause is still unclear. The data of other installations, such as KASCADE-Grande (Chiavassa et al 2019), IceTop (Aartsen et al 2016), do not cover this region, but confirm the trends in the energy dependences of the anisotropy amplitude and phase.

In this paper, we present a method of anisotropy search based on muon bundle detection, as well as a method of atmospheric effect correction. In contrast to other experiments, we use the local muon density measured in the detector for primary energy estimating.

### 3. EXPERIMENTAL SETUP

To investigate cosmic rays in a wide range of primary energies, the Experimental complex NEVOD (Petrukhin 2015) has been constructed in MEPhI (Moscow, Russia) at the ground surface, about 164 m above the sea level. The complex includes several scientific installations for EAS components investigations (Yashin et al 2021). One of them is the coordinate-tracking detector DECOR (Barbashina et al. 2000). It includes 8 supermodules (SM) with a total area of about 70 m$^2$ which are arranged around the 2000 m$^3$ water tank of the Cherenkov water detector NEVOD, as shown in Figures 1 and 2. Two supermodules are installed in each of two short galleries of the laboratory building (SMs 0, 1, 6, 7) and four supermodules (SMs 2, 3, 4, 5) are located in the long one. The main DECOR features are the vertical deployment of detecting planes and good spatial and angular resolutions (~ 1 cm and ~ 1°) for inclined muon tracks. Each supermodule represents an eight-layer system of plastic streamer tubes with a resistive coating of the cathode. The planes are suspended at a distance of 6 cm from each other. Each layer is equipped with aluminum strips forming two-coordinate readout system (vertical and horizontal).

The muon bundle is a group of several genetically related muons with quasi-parallel tracks. Muon bundles are generated close to the air-shower core and can reach the detector in a wide range of zenith angles. Moreover, muon bundles retain the direction of primary particle with a good accuracy. Therefore they can be used to study CR anisotropy.

The selection of events is carried out in several steps. Firstly, at the trigger system level, at least three DECOR supermodules must be hit within the time gate of 250 ns. Then, software reconstruction is used to select events by the number of quasi-parallel (within a five-degree cone) tracks. An example of a muon bundle event detected with the coordinate-tracking detector is shown in Figures 1 and 2. Finally, we select events in which at least 3 muon tracks with arrival direction zenith angles in the range from 15° to 75° are detected in at least 3 different SMs.

As mentioned above, the DECOR supermodules have a vertical orientation. This provides a reliable detection of near-horizontal particle tracks. At small zenith angles, the effective area of the detector decreases, and the background from the electron-photon component increases. Therefore, the range of zenith angles from 0° to 15° was excluded from the analysis. On the other hand, the intensity of muon bundles rapidly decreases with the growth of zenith angles. In such conditions, the relative fraction of events with incorrect geometry reconstruction increases at large zenith angles. That is why the events with zenith angles greater than 75° were also excluded.

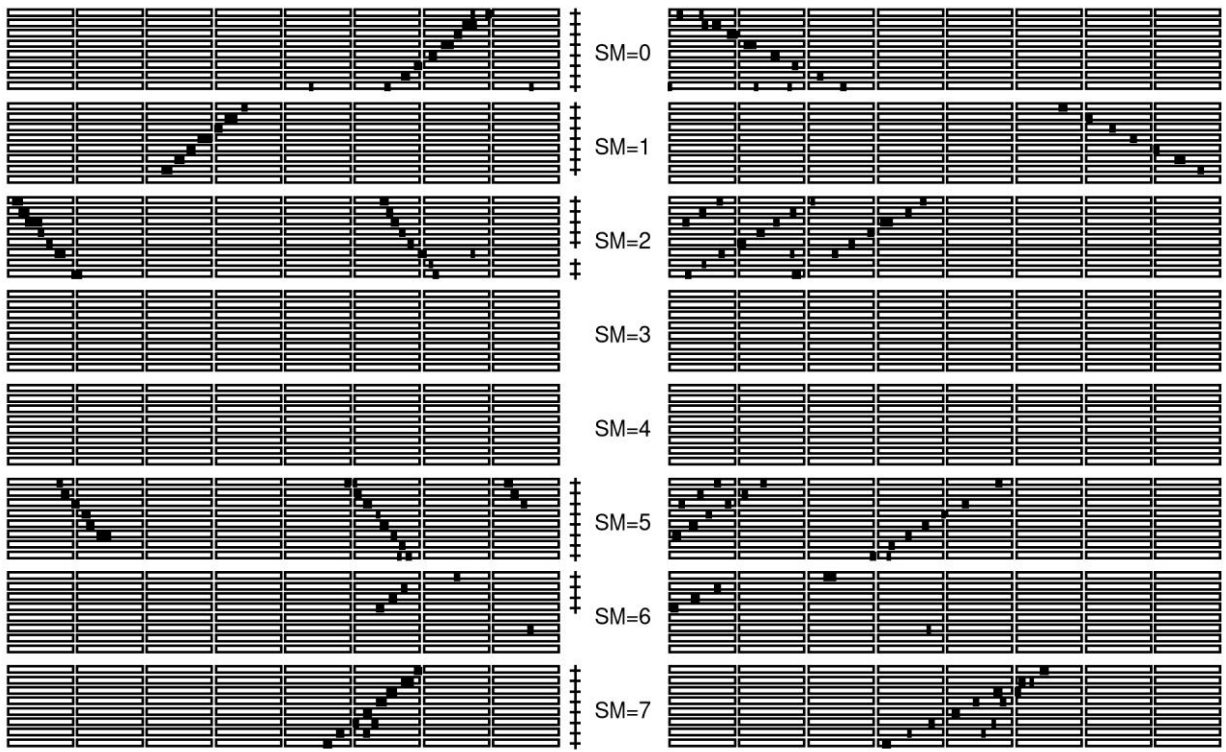

**Figure 1.** Response of DECOR supermodules to a muon bundle event.

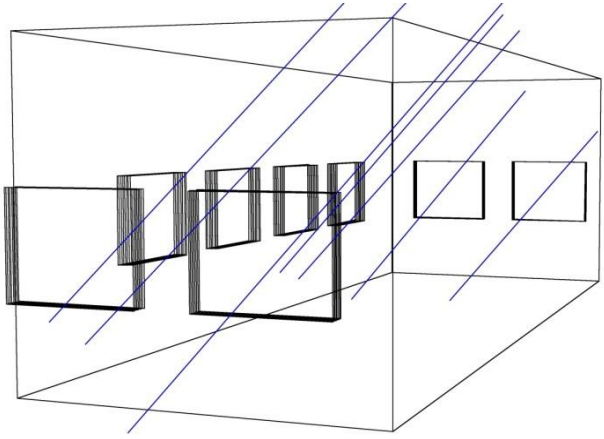

**Figure 2.** Geometrical reconstruction of the muon bundle event. Big rectangles are DECOR supermodules.

## 4. PRIMARY ENERGY ESTIMATION

Estimation of the energy of primary particles was carried out with simulated events using the spectra of local muon density measured for different zenith angles (Bogdanov et al. 2010). Muon bundles were simulated in the range of zenith angles from 15° to 75° and azimuthal angles from 0° to 90°. The trigger conditions for selecting simulated events were the same as for real events (see section 3). The "equivalent time" of a set of simulated events corresponds to about 254 hours of experimental data taking.

To compare the obtained sample of simulated events with the experiment, we used a set of experimental runs with duration of 3459 hours ("live time"). During this time, 830.8 thousand events with arrival directions in the range of zenith angles from 15° to 75° were recorded. As seen in Figure 3, angular distributions for simulated and experimental events are in very good agreement. To compare the distributions over the azimuth angle, the experimental data were summarized over 4 quadrants; the total number of detected events was normalized to the number of simulated ones.

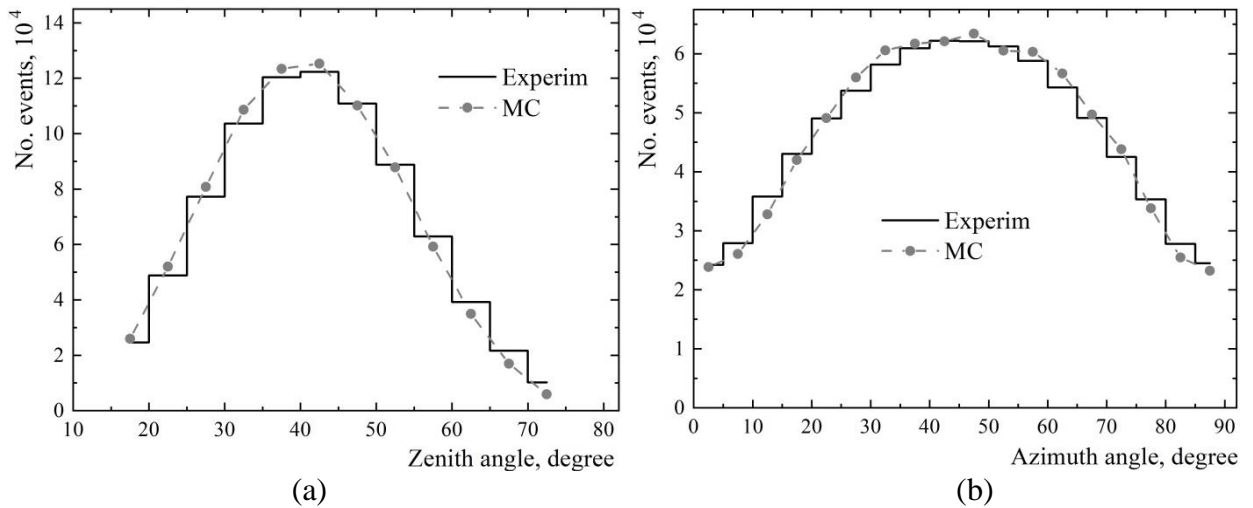

(a)        (b)

Figure 3. Comparison of distributions in zenith (a) and azimuthal (b) angles.

To determine the density of muons in a bundle, the effective area of the detector, which depends on the particle arrival direction, was taken into account. To pass from muon density distributions to primary particle energy distributions, it is necessary to use the estimative formula (Bogdanov et al. 2010), which relates the average logarithm of the primary energy to the muon density and zenith angle of the particle arrival direction:

$$\langle \lg(E_{est}, \text{GeV}) \rangle \approx 7.03 + 1.07 \, lg(D, \text{m}^{-2}) + 3.80 \, lg(\sec\theta) \quad (5)$$

The variable $E_{est}$ as an intermediate estimator of the primary energy. The distributions of the logarithm of the primary energy estimator ($\lg E_{est}$) obtained using formula (5) in the simulated events for muon track multiplicities $m \geq 3$ and $m \geq 5$ are shown in Figure 4. The distribution for the minimal multiplicity of muon tracks $m = 3$ is rather wide. The long left tail is due to the Poisson fluctuations in the number of tracks in events with low muon density. With the increase of the threshold number of tracks (transition from $m = 3$ to $m = 5$) the distribution becomes narrow, but the statistics decreases by more than 4 times. Numerical characteristics of distributions of the muon bundle density $D$ and the logarithm of the primary energy estimator $\lg E_{est}$ (mean values and root-mean-square deviations) are given in Table 1.

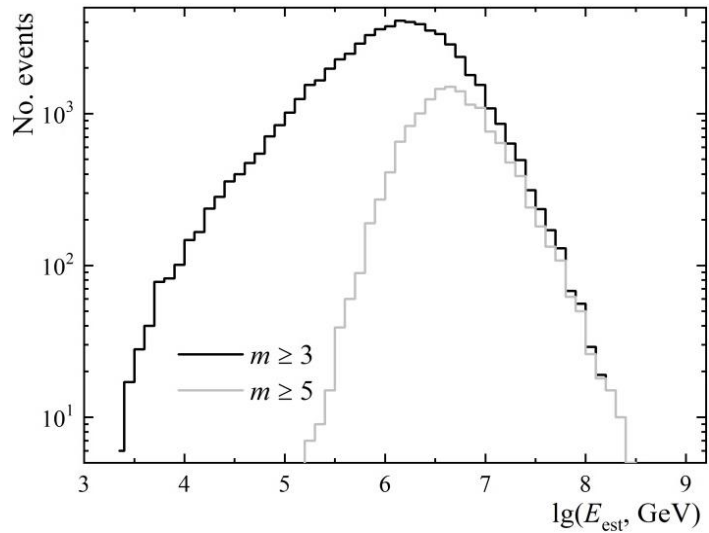

Figure 4. Distributions of the logarithm of the primary energy estimator (5) for simulated events.

**Table 1.**

Characteristics of distributions for two samples of simulated events.

| $m$ | no. ev. | $\langle \lg(D, \text{m}^{-2}) \rangle$ | $\sigma(\lg D)$ | $\langle \lg(E_{est}, \text{GeV}) \rangle$ | $\sigma(\lg E_{est})$ | $\langle \lg E_0, \text{GeV} \rangle$ | $\sigma(\lg E_0)$ |
|---|---|---|---|---|---|---|---|
| $\geq 3$ | 61831 | $-1.418 \pm 0.002$ | 0.616 | $6.057 \pm 0.003$ | 0.689 | $6.06 \pm 0.10$ | 0.80 |
| $\geq 5$ | 14555 | $-0.820 \pm 0.003$ | 0.369 | $6.690 \pm 0.004$ | 0.439 | $6.69 \pm 0.10$ | 0.59 |

For a correct assessment of the parameters of distribution in logarithm of the primary energy ($lgE_0$), it should be taken into account that formula (5) is valid only for the average values of $D$ and $lg(sec\theta)$. The uncertainty of the average value of $lgE_0$ is 0.10 (Bogdanov et al. 2010) and is associated not only with the approximation of formula (5), but also with the choice of the model of hadronic interaction, the assumption on CR composition, etc.

The root-mean-square spread of the logarithm of the primary particle energies, corresponding to a fixed muon density and a fixed arrival direction, is about 0.4 (Bogdanov et al. 2010). Such spread is associated with the randomness of the EAS core positions relative to detector location, and its estimate weakly depends on the angle, density, model, and type of primary nuclei. To account for this source of spread, it necessary to add (in squares) 0.4 to the earlier obtained $\sigma(lgE_{est})$. The result of this operation is shown in the last column of Table 1. Thus, two samples of data with minimal track multiplicities ($m \geq 3$ and $m \geq 5$) correspond to mean logarithmic energies of primary particles of about 1 PeV and 5 PeV, respectively.

In the Galaxy, the gyroradius for protons with such energies is about 1 – 5 pc or 3 – 15 light years. It is comparable with the distances to the nearest stars. In this region of space, no particle sources producing particles with such high energies have been discovered yet. So, it is possible to study large-scale anisotropy, while the search for SSA is rather complicated due to the rapid decrease of intensity with the growth of primary energy.

## 5. EXPERIMENTAL DATA

The experimental data for the analysis of CR anisotropy were collected in four continuous series of data taking (see Table 2) performed in 2012–2022 at the DECOR detector. The total "live time" of observation was 57 271 hours (~ 2 386 days). The entire period of data taking is divided into four rather long intervals called "series". This division is associated with some minor changes in the configurations of the installations included in the experimental complex. Within each series, the data are segmented into time intervals (called "runs") lasting from 10 to 40 hours. Events with muon bundles were selected according to conditions described in section 3. The durations of the series and the numbers of events recorded during each one for two muon bundle multiplicity thresholds are given in the Table 2. With an increase in the threshold from 3 to 5 tracks the number of events decreases by about 3.4 times.

In the local coordinate system of the experimental complex, the distribution of events in the zenith angle does not practically change its shape when passing from one threshold to another (Figure 5, left). The distributions in the azimuthal angle (Figure 5, right) indicate 4 preferred directions of detection associated with the azimuthal dependence of the effective area of the DECOR detector.

**Table 2.**

The series of data taking.

| Series | Start | End | No. events ≥ 3 tracks | No. events ≥ 5 tracks |
|--------|-------|-----|----------------------|----------------------|
| 10 | 2012/05/03 | 2013/03/02 | 1 188 690 | 355 777 |
| 11 | 2013/06/04 | 2015/04/08 | 2 922 316 | 874 525 |
| 12 | 2015/07/16 | 2018/12/25 | 5 878 201 | 1 760 296 |
| 13 | 2019/09/18 | 2022/01/19 | 3 956 759 | 1 166 203 |
| Total | 2012/05/03 | 2022/01/19 | 13 945 966 | 4 156 801 |

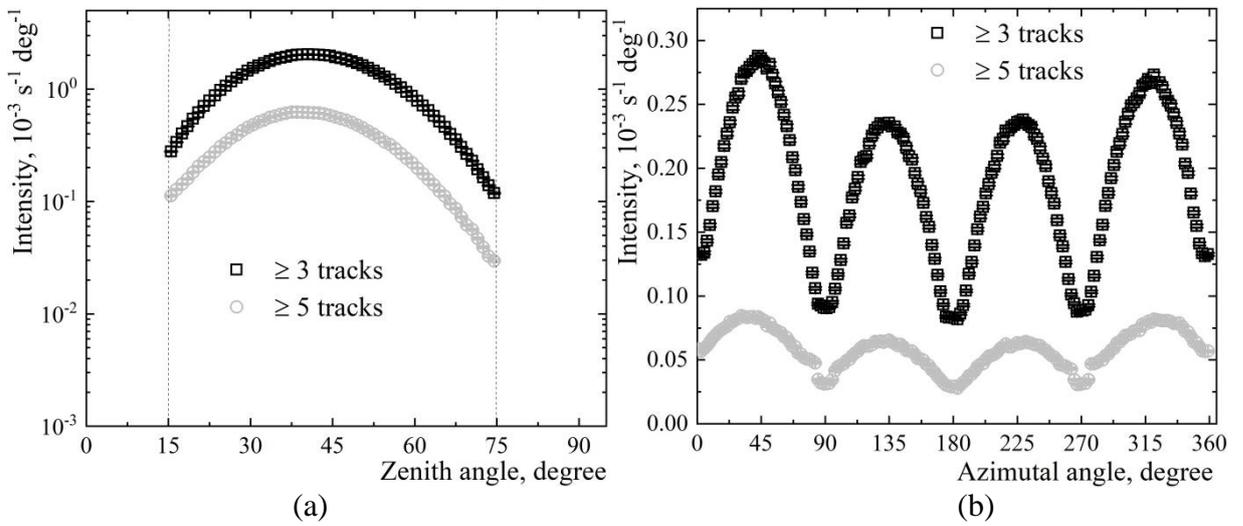

**Figure 5.** Distributions of events in zenith (a) and azimuthal (b) angles.

## 6. CORRECTION FOR ATMOSPHERIC EFFECT

The muons are generated in the atmosphere mainly through decays of charged pions and kaons which are components of nuclear cascades initiated by high energy primary cosmic ray particles. This air shower process develops in the atmosphere and, consequently, the parameters of the medium may affect the muon density at the altitude of the detector location. Long-term measurements of the muon bundles with DECOR detector have revealed variations in the intensity of muon bundles (Figure 6, grey squares). Each point in the plot represents the average rate for one run of a series. The blank spaces in Figure 6 separate data of different series. There are both seasonal and short-term variations in the counting rate of the muon bundles.

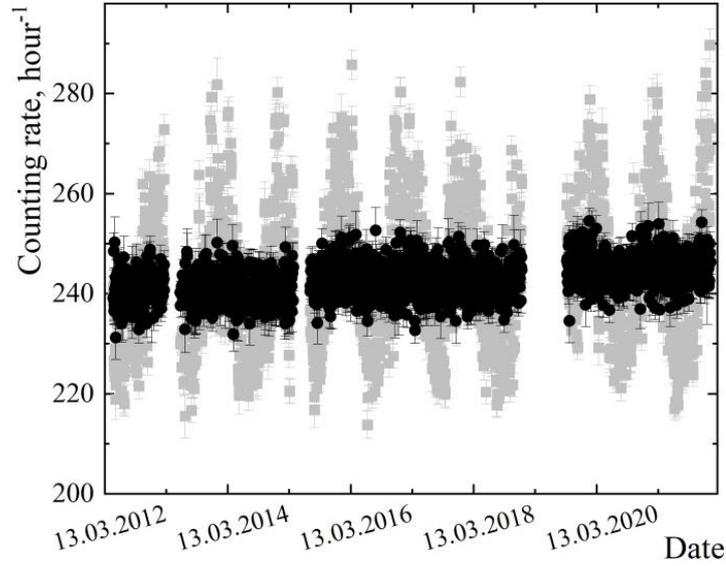

**Figure 6.** Variations in the muon bundle counting rate (grey squares). The data corrected for atmospheric effect are represented by black circles.

The muon bundle multiplicity is related to the parameters of the lateral distribution function (LDF) of muons at the detection level, which in its turn depends on the altitude of their generation in the atmosphere. With the decrease of the altitude of generation point, the amount of detected events satisfying the selection conditions increases. Based on the analysis of variations of muon bundle counting rate, the method for atmospheric effect compensation (Yurina et al. 2019) was developed in the Experimental complex NEVOD. This method uses close correlations (correlation coefficient of 0.97) of the muon bundle counting rate with the altitude of the atmospheric layer with residual pressure of 500 mbar ($H_{500}$). The best approximation of relation between the counting rate and the altitude ($H_{500}$) is achieved with using of a power-low function:

$$F(t) = F_0 \left(\frac{H_{500}(t)}{H_0}\right)^\lambda = \eta(t) F_0, \qquad (6)$$

where $t$ is the date and time of measurement, $H_0$ and $F_0$ are the altitude of atmospheric layer with the residual pressure of 500 mbar and the muon bundle counting rate averaged over a long period of time, respectively, power index $\lambda$ is related to the slope of the spectrum of local muon densities (Yurina et al. 2019). Data on $H_{500}$ values were taken from the Global Data Assimilation System (GDAS) database (NOAA 2022) for the node closest to the detector (37°N, 56°E) in a one-degree geographic grid. The estimates of parameters of function (6) for each series are presented in Table 3. During more than 9 years of operation, the configuration of the detector has been slightly changed between the series. The changes were mainly concerned with the adjustment of thresholds for recording pulses from streamer tubes, and with some background conditions. So, the average counting rates in series are slightly different (see Table 3).

**Table 3.**

The parameters of function (6) and average counting rates of muon bundles for each series of data taking.

| Series | $H_0$, m | $\lambda$ | Rate, h$^{-1}$ | Corrected Rate, h$^{-1}$ |
|---|---|---|---|---|
| 10 | 5524 ± 12 | -1.684 ± 0.034 | 240.9 ± 12.1 | 240.8 ± 3.2 |
| 11 | 5541 ± 9 | -1.688 ± 0.020 | 241.8 ± 13.5 | 240.8 ± 3.2 |
| 12 | 5532 ± 7 | -1.674 ± 0.015 | 243.6 ± 13.9 | 242.9 ± 3.4 |
| 13 | 5515 ± 10 | -1.659 ± 0.021 | 246.3 ± 14.6 | 245.0 ± 3.2 |

The rate errors indicated in two last columns of Table 3 represent the run-to-run r.m.s spread. The correction for atmospheric effect provided decreasing of the r.m.s spread in about 4 times (Figure 6, black points), and its value (Table 3) almost reached the level of statistical errors of about 1 – 2 %. This fact demonstrates that the seasonal effect and short-term modulations of atmospheric origin were practically eliminated.

## 7. ESTIMATION OF THE EXPECTED NUMBER OF EVENTS

According to formula (3), the amplitude of CR flux anisotropy is defined as a relative deviation of the number of recorded events ($N_{\text{rec}}$) from the amount expected in assumption of isotropic flux along the selected direction in a given solid angle ($N_{\text{exp}}$):

$$A = \frac{N_{\text{rec}} - N_{\text{exp}}}{N_{\text{exp}}}. \tag{7}$$

To estimate the error ($\Delta A$) of this quantity ($A$), the following expression is used:

$$\Delta A = \frac{\Delta N_{\text{rec}}}{N_{\text{exp}}} = \frac{\sqrt{N_{\text{rec}}}}{N_{\text{exp}}}. \tag{8}$$

The most convenient way to obtain the expected number of events from a given detection direction is to use the 2$^{\text{nd}}$ equatorial coordinate system (Sadler et al. 1974), because this system and the Earth have a common equatorial plane. As an alternative, the galactic coordinate system (Blaauw et al. 1960) may be used. Both systems use two spherical coordinates: latitude and longitude similar to those which are used on the Earth. To present the measurement results we used the HEALPix framework (Gorski et al. 2005) with Hammer-Aitoff equal-area projection (Calabretta M.R. & Greisen E.W. 2002) for the pixelization of data on the celestial sphere. The HEALPix grid resolution parameter $N_{\text{side}}$ is set to 64 for all pictures. It corresponds to sell size of about 2.6×10$^{-4}$ sr on the celestial sphere.

When estimating the expected number of events, it is important to take into account the factors affecting the detection of muon bundles. These factors include: background conditions, design features that affect the detection efficiency for different arrival directions of muon bundles, atmospheric effects and the inhomogeneity of time of observation of various parts of the celestial sphere. The estimation method is based on the usage of the available sample of recorded events. To exclude the influence of different background conditions when changing the detector configuration (see Section 5), we processed the data of each series separately and then merged the results. To compensate atmospheric effect, we summed the values inverse to the coefficients η(*t*) in formula (6) for each cell of the celestial sphere:

$$N(\alpha, \delta) = \sum \eta^{-1}(t_i), \qquad (9)$$

where (α, δ) are coordinates of the cell on the celestial sphere, and *i* is the event number in the given cell. To take into account the design features we used so called "aperture snapshot" (Figure 7). These 2D-distributions in the 2$^{nd}$ equatorial and galactic coordinate systems are collected assuming that the sidereal time is constant (for example, it equals to zero). In other words, we virtually "stopped" the Earth's rotation around its axis. In this case, the snapshot of all events is the image of the muon bundle flux multiplied by the aperture of the DECOR detector. In Figure 7, colors indicate the number of events detected in each cell. Four colored spots on the figure correspond to four highlighted directions in the local coordinate system (see Figure 5). The white spot corresponds to directions with near-vertical zenith angles (θ < 15°). Its center corresponds to the zero zenith angle in the local coordinate system.

In the case of an isotropic flux, all directions are considered as equal, and the number of recorded events in a given solid angle is only related to the duration of measurement. So, to obtain the map of the expected numbers of events, we rotate the aperture snapshot around the world axis together with the Earth and normalize it to the "live time" of observation.

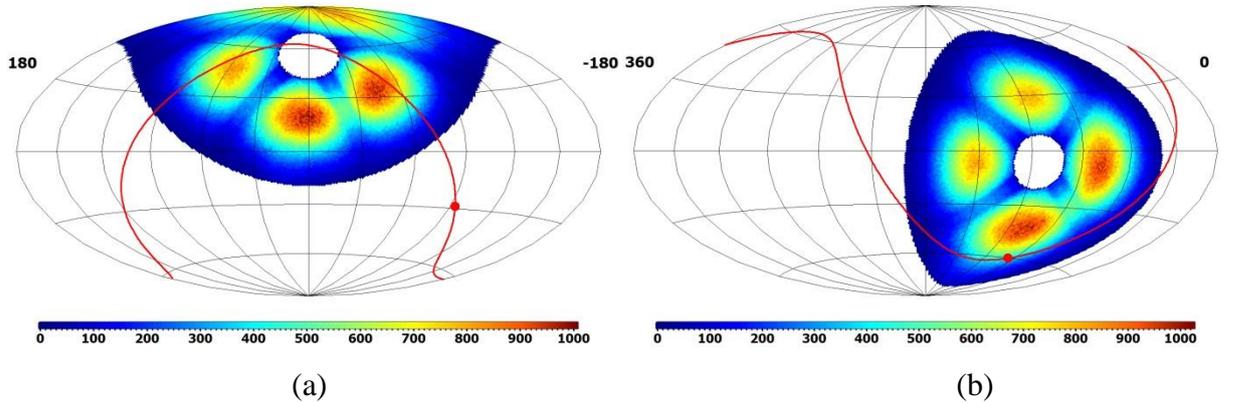

(a)  (b)

**Figure 7.** Distribution of all events from the sample of 12$^{th}$ series in the celestial sphere for the zero moment of sidereal time plotted in the 2$^{nd}$ equatorial (a) and in the galactic (b) coordinate systems. Red line represents the Galaxy plane (a) and equatorial plane (b). The red point shows the center of Galaxy (a) and the reference point of the 2$^{nd}$ equatorial coordinate system (b).

The "live time" of observation is the part of detector operation time after subtraction of the time required for hardware and software processing of events. The distribution of "live time" in UTC system (solar time) has a dip in the morning period of the day (see Figure 8, left). Its main reason is interruptions of exposition for the detector maintenance. The standard deviation from the mean value of 2386.3 days is approximately 3.4 %. The distribution of "live time" within the sidereal day for the entire period of observation is almost uniform (see Figure 8, right), the standard deviation from the average value of 2392.8 days is less than 0.6%.

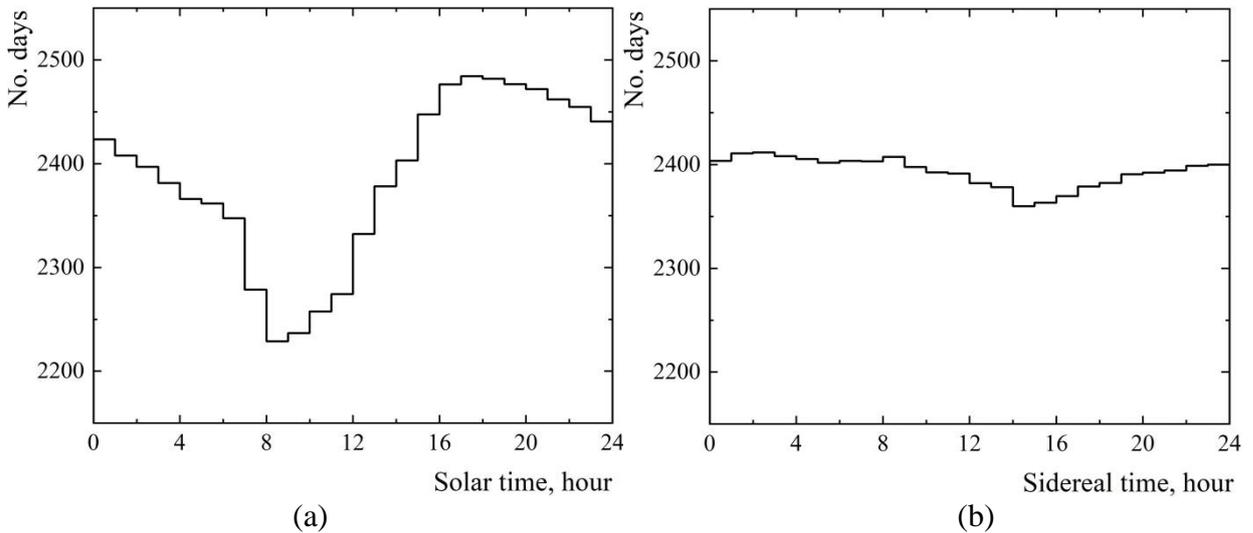

(a)          (b)

**Figure 8.** The distribution of "live time" within the solar (left) and sidereal (right) day.

As a result, the distribution of the expected number of events over the celestial sphere has near-axial symmetry (Figure 9) in the $2^{nd}$ equatorial coordinate system and is uneven over the declination. Significant part of events is detected from the polar region in the $2^{nd}$ equatorial coordinate system. The expected number of events decreases as it approaches the equatorial plane and disappears in the southern celestial hemisphere. The total declination range is from -20° to 90°. That is why we cannot observe the area of the celestial sphere where the Galactic center is located ($\alpha = 266°$, $\delta = -29°$, see Figure 9). The same image plotted in the galactic coordinate system demonstrates that we observe mainly the outer region of the Galaxy, but the data are available over the entire range of galactic longitudes.

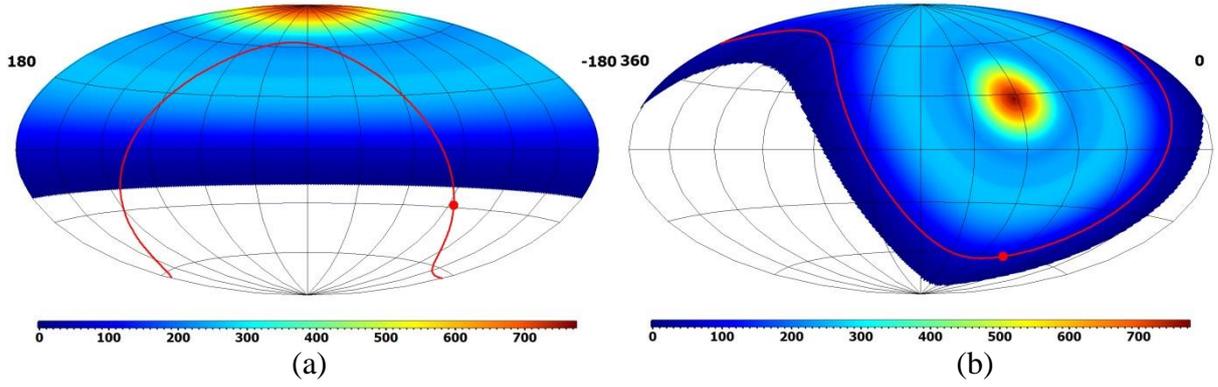

**Figure 9.** Distribution of the expected number of events with muon bundles from the sample of the 12[th] series plotted in the 2[nd] equatorial (left) and in the galactic (right) coordinate systems. Red line represents the Galaxy plane (left) and equatorial plane (right). The red point shows the center of the Galaxy (left) and the reference point of the 2[nd] equatorial coordinate system (right).

## 8. RESULTS AND DISCUSSION

Unfortunately, the collected statistics of events still does not allow us to obtain a clear image of the anisotropy of the cosmic ray flux on the celestial sphere. The results of data processing for two threshold values of the number of registered tracks are presented in one-dimensional projections on the right ascension and galactic longitude (see Figures 10 and 11). The relation between the minimal number of recorded tracks in the sample and the average logarithm of the primary particle energy is described in Section 4. To plot the dependences, the entire ranges of right ascension and galactic longitude were divided into 18 bins of 20° degrees each. The data of each series of data taking were analyzed independently from each other. The result for the entire period was obtained as a weighted average, where the weights were determined by the respective statistical errors.

In accordance with (4), we approximated these dependences by the function:

$$f(\alpha) = b + a\cos(\alpha - \varphi). \tag{10}$$

The results of fitting are presented in Table 4. In case of the 2[nd] equatorial coordinate system, the parameter $b$ on the left plots can be considered as equal to zero. In each case, the normalized value of $\chi^2$/dof (dof indicates here the number of degrees of freedom) is higher for the hypothesis of isotropic CR flux. It means that the option of the presence of dipole anisotropy is preferable. The amplitude of the dipole anisotropy corresponds to a significance of about 2.6 σ in both the 2[nd] equatorial and the galactic coordinate systems. The value of the anisotropy phase determined in both coordinate systems agrees with the position of the Galactic center. The deviations of the obtained anisotropy phases in the 2[nd] equatorial ($\Delta\varphi_{Eq}$) and in the galactic ($\Delta\varphi_{Gal}$) coordinate systems from the direction to the Galactic center ($\alpha_{GC}=266°$) are given in Table 4.

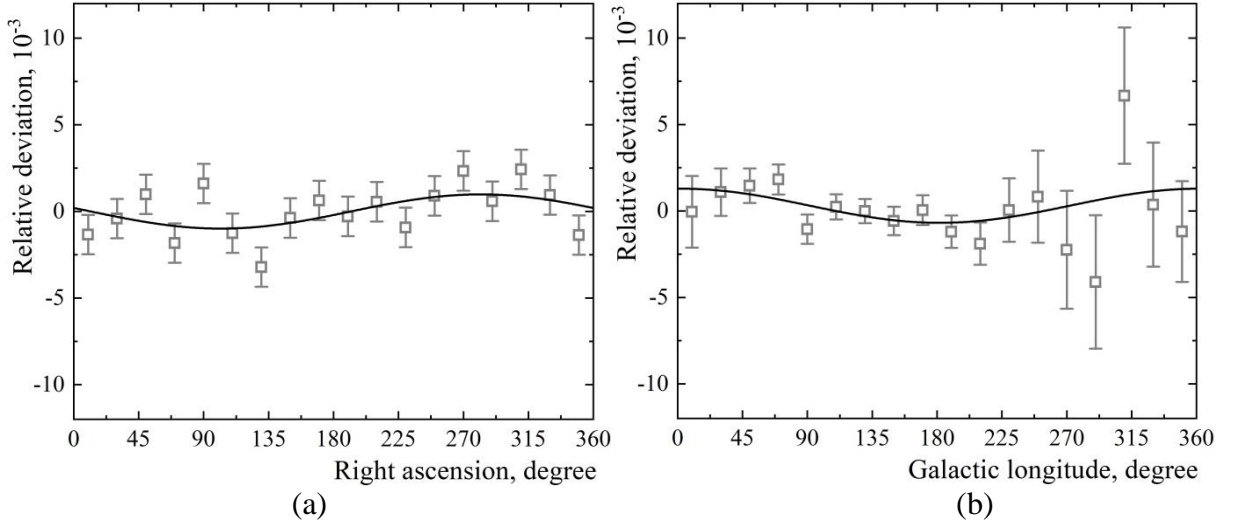

**Figure 10.** One-dimensional projections of the relative deviations for selection threshold of 3 muon tracks on the right ascension (a) and on the galactic longitude (b). The sinusoidal curves represent the best fits.

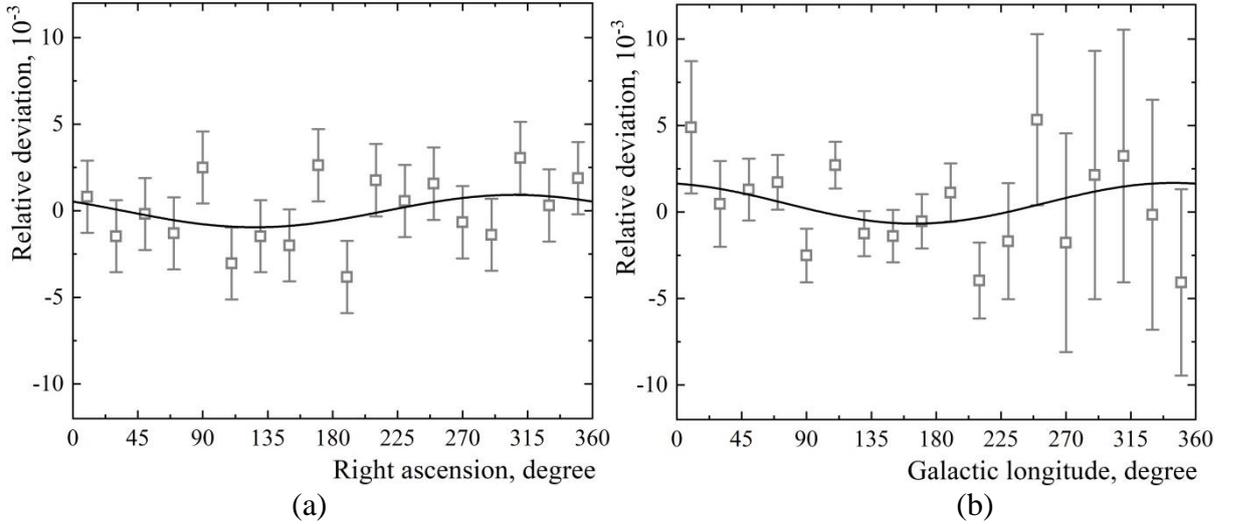

**Figure 11.** One-dimensional projections of the relative deviations for threshold of 5 muon tracks on the right ascension (a) and on the galactic longitude (b). The sinusoidal curves represent the best fits.

The results of dipole anisotropy measurements in the energy range of primary particles from 0.1 to 30 PeV obtained in different experiments are presented in Figure 12. The anisotropy phase drastically changes near the energy of 0.2 PeV. At energies from few hundred TeV up to about 30 PeV, the anisotropy phase is almost constant. Also, it can be concluded that the phase coincides with the direction to the center of the Galaxy. The dashed line in the upper plot of Figure 12 shows a graph of the dependence of the anisotropy amplitude on the primary energy assumed by diffusion theory (Berezinsky 1990).

**Table 4**

The parameters of dipole anisotropy approximation

| System | <lg(E,TeV)> | $b$, $10^{-3}$ | $a$, $10^{-3}$ | $\Delta\varphi_{Eq}/\Delta\varphi_{Gal}$ (degree) | $\chi^2/dof^a$ (dipole) | $\chi^2/dof$ (isotropic) |
|---|---|---|---|---|---|---|
| 2nd Equat. | 3.06 ± 0.80 | 0.00 ± 0.27 | 1.01 ± 0.38 | 16 ± 22 | 22.2/15 | 43.3/15 |
| 2nd Equat. | 3.69 ± 0.59 | -0.02 ± 0.49 | 0.95 ± 0.69 | 40 ± 42 | 14.0/15 | 19.8/15 |
| Galactic | 3.06 ± 0.80 | 0.30 ± 0.27 | 1.01 ± 0.39 | 2 ± 21 | 12.6/15 | 32.3/15 |
| Galactic | 3.69 ± 0.59 | 0.51 ± 0.49 | 1.20 ± 0.69 | -15 ± 33 | 16.1/15 | 24.1/15 |

[a]dof indicates the number of degrees of freedom

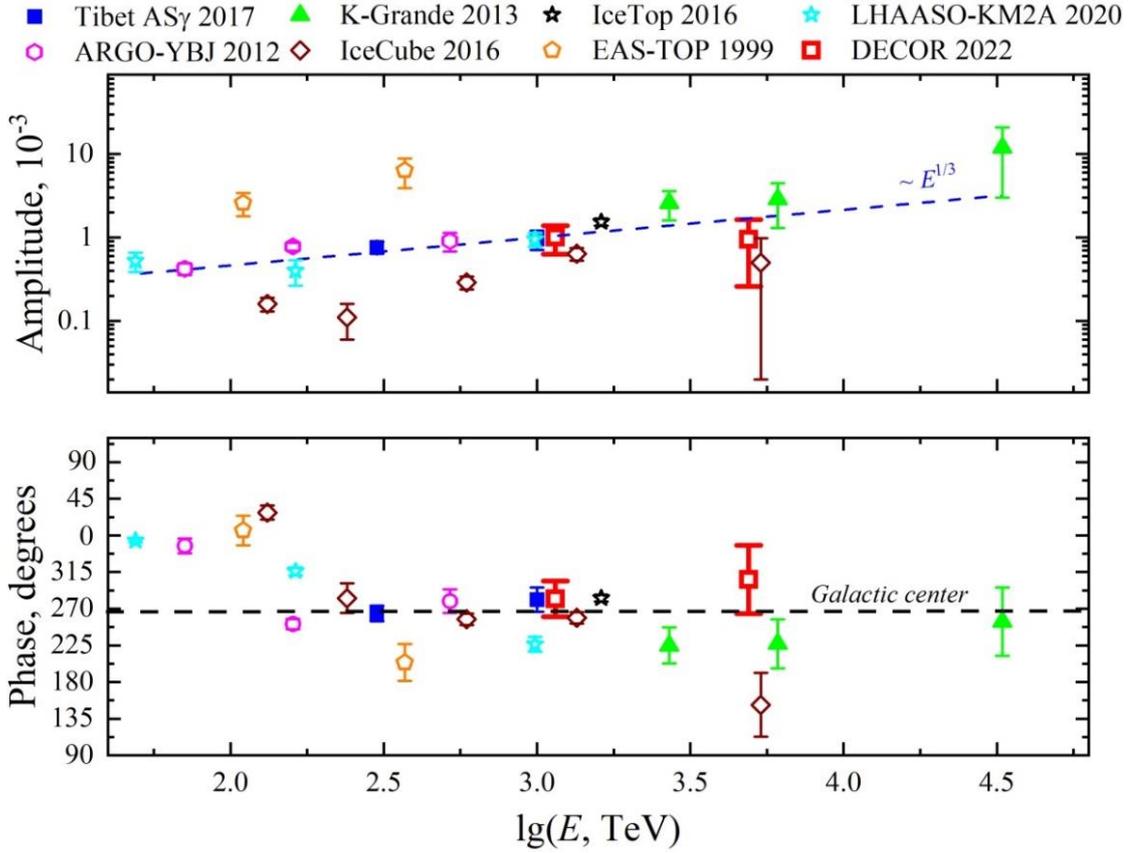

**Figure 12.** Parameters of dipole anisotropy obtained in different experiments.

Our results confirm the presence of dipole anisotropy with a significance of 2.6 σ. It is possible that with higher statistics we will be able to increase this value. Currently, the parameters of the dipole anisotropy obtained in our research using new methods for anisotropy measurements are consistent with the data of such installations as IceCube (Aartsen et al. 2016) and Tibet ASγ array (Amenomori et al. 2017) (see Figure 12).

# 9. CONCLUSIONS

The research results presented in this paper indicate that muon bundles can be used as a tool for studying the anisotropy of cosmic rays. The methods of data recording and processing developed at the Experimental Complex NEVOD provide suppression and accounting for hampering factors, as well as obtaining of reliable results.

It should be mentioned that cosmic rays anisotropy has never been studied by means of muon bundles. Such measurements have been performed for the first time with the coordinate-tracking detector DECOR. Also at the Experimental complex NEVOD, the new method of correction for atmospheric effects was developed and implemented. The advantage of this method is that the height of the atmospheric layer with a residual pressure of 500 mbar is the only parameter necessary for the correction. The results of the performed data analysis demonstrate that the developed method provides almost complete suppressing of the influence of atmospheric effects on the counting rate of muon bundles.

The effective energy of primary particles is related to the density of the detected muon bundles. The average logarithmic energies of primary particles are about 1 PeV and 5 PeV for the selection conditions of at least 3 and 5 muon tracks, correspondingly.

The method for estimating the number of events under the assumption of an isotropic flux, which we used in our research, provides an opportunity to take into account the design features of the detector and the non-uniformity of its detection efficiency for different arrival directions of muon bundles.

The obtained dipole anisotropy parameters are consistent with the theory of diffuse propagation of CR from the center of the Galaxy.

## Acknowledgements

The work was performed at the Unique Scientific Facility "Experimental complex NEVOD" with the financial support provided by the Russian Ministry of Science and Higher Education, project "Fundamental problems of cosmic rays and dark matter", No. 0723-2020-0040.

## References

Aartsen, M.G., Abraham, K., Ackermann, M., et al. 2016, ApJ, 826, 220

Amenomori, M., Bao, Y.-W., Bi, X. J. et al. 2019, PoS(ICRC2019)488

Barbashina, N.S., Ezubchenko, A.A., Kokoulin, R.P., et al. 2000, IET, 43, 743

Bartoli, B., Bernardini, P., Bi, X.J., et al. 2013, PhRvD, 88, 082001

Blaauw, A., Gum, C.S., Pawsey, J.L., et al. 1960, MNRAS, 121(2), 123.

Berezinsky, V. S. 1990, 21st ICRC, Cosmic ray propagation in the Galaxy (Adelaide), 11, 115

Bogdanov, A.G., Gromushkin, D.M., Kokoulin, R.P., et al. 2010, PAN, 73, 1852

Bonino, R., Alekseenko, V.V., Deligny, O., et al. 2011, ApJ, 738, 67

Calabretta, M.R., & Greisen, E.W. 2002, A&A, 395, 1077

Chiavassa, A., Apel, W.D., Arteaga-Velázquez, J.C., et al. 2019, ApJ, 870, 91

Compton, A.H., & Getting, I.A. 1935, PhRv, 47, 817

Gao, W., Cao, Q., Chen, S., et al. 2021, PoS(ICRC2021) 351

Gleeson, L.J., & Axford, W.I. 1968, Ap&SS, 2, 431

Gorski, K.M., Hivon, E., Banday, A.J., et al. 2005, ApJ, 622, 759

Longair M.S. 2011, High Energy Astrophysics, Vol. 2, 377 (3rd ed.; Cambridge University Press) doi.org/10.1017/CBO9780511778346

NOAA Air Resources Laboratory (ARL) 2022, Global Data Assimilation System (GDAS) Archive Information, http:/ready.arl.noaa.gov/gdas1.php

Petrukhin, A.A. 2015, PhyU, 58, 486

Reid, M.J., & Zheng, X.-W. 2020, SciAm, 322, 4:28

Reid, M.J., Menten, K.M., Brunthaler, A., et al. 2014, ApJ, 783, 130

Sadler, D.H., McBain Sadler, F.M., Porter, J.G., et al. 1974, Explanatory Supplement to the Astronomical Ephemeris and the American Ephemeris and Nautical Almanac (3rd ed.; London, issued by Her Majesty's stationery office)

Schönrich, R., Binney, J., & Dehnen W. 2010, MNRAS, 403, 1829

Yashin, I.I., Amelchakov, M.B., Astapov, I.I., et al, 2021, JInst, 16, T08014

Yurina, E.A., Bogdanov, A.G., Dmitrieva, A.N., et al. 2019, Journal Physics: Conference Series, 1189, 012010